\documentclass[floatfix,twocolumn,showpacs,prb]{revtex4}
\usepackage{graphicx}
\usepackage{amsmath}
\usepackage{bm}
\usepackage{dcolumn}
\usepackage{color}

\begin{document}

\title{Non-collinear magnetism in freestanding and supported monatomic Mn chains}

\author{Franziska Schubert$^{1}$}
\altaffiliation[Permanent address: ]
{Fritz-Haber-Institute der Max-Planck-Gesellschaft, Berlin, Germany}
\author{Yuriy Mokrousov$^{2}$}
\email{y.mokrousov@fz-juelich.de}
\author{Paolo Ferriani$^{3}$}
\author{Stefan~Heinze$^{3}$}

\affiliation{$^1$Institute of Applied Physics,
University of Hamburg, D-20355 Hamburg, Germany}
\affiliation{$^2$Institute for Advanced Simulation, Forschungszentrum J\"ulich and JARA,
D-52425 J\"ulich, Germany}
\affiliation{$^3$Institute of Theoretical Physics and Astrophysics,
Christian-Albrechts-University of Kiel, D-24098 Kiel, Germany}

\date{\today}

\begin{abstract}

Using first-principles calculations, we study the occurrence of non-collinear magnetic
order in monatomic Mn chains. First, we focus on freestanding Mn chains and demonstrate
that they exhibit a pronounced non-collinear ground state in a large range of interatomic
distances between atoms in the chain. By artificially varying the atomic number of Mn we
investigate how the magnetic ground state is influenced by alloying the Mn chains with 
Fe and Cr. With increasing number of $3d$-electrons we find a smooth transition in the
magnetic phase space starting from an antiferromagnetic state for pure Cr chains through 
a regime of non-collinear ground states for Mn-rich chains to a ferromagnetic solution 
approaching the limit of pure Fe chains. Second, we investigate the magnetism in 
supported Mn chains on the (110)-surfaces of Cu, Pd, and Ag. We show that even a weak 
chain-surface hybridization is sufficient to dramatically change the magnetic coupling in 
the chain. Nevertheless, while we observe that Mn chains are antiferromagnetic on 
Pd(110), a weak non-collinear magnetic order survives for Mn chains on Cu(110) 
and Ag(110) a few meV in energy below the antiferromagnetic solution. 
We explain the sensitive dependence of the exchange interaction in Mn 
chains on the interatomic distance, chemical composition, and their 
environment based on the competition between the ferromagnetic double 
exchange and the antiferromagnetic kinetic exchange mechanism. 
Finally, we perform simulations which predict that the non-collinear 
magnetic order of Mn chains on Cu(110) and Ag(110) could 
be experimentally verified by spin-polarized scanning 
tunneling microscopy.
\end{abstract}


\maketitle

\section{Introduction}

In pursuit of atomic-scale magnetic storage devices and future spintronics applications,
one-dimensional (1D) transition-metal (TM) nanostructures have become a topic 
of intense research interest.\cite{Gambardella,Mn_CuN_Heinrich,Calvo2009,Serrate}
A promising route for future applications concerns e.g.~the control of the spin state in
1D wires using spin-polarized currents via the spin transfer torque.
Besides the strong technological motivation, from the fundamental point of view 
one-dimensional systems constitute a unique playground for electronic structure theory. 
Due to enhanced intra-atomic exchange and lowered coordination, they reveal an 
increased tendency towards magnetism. For example, elements such as Pt and Pd, which 
are non-magnetic in bulk, can become magnetic in atomic chains.\cite{Pt-Smog,
Pd-Tosatti} Moreover, giant values of the magnetic anisotropy energy, which is crucial 
to stabilize magnetism against thermal fluctuations, have been found for free-standing, 
suspended, and deposited TM chains.\cite{Gambardella,Giant4d,Pt-Smog} 
Owing to the rich physics in one-dimensional systems many more novel effects and
phenomena have been reported in the recent years concerning their
electronic structure, spin dynamics and transport properties.
\cite{Tsymbal2007,Samir2008,MokrousovFeIr,
Maz-NM,Calvo2009,Abanov2010,Bauer2010}

The creation of monatomic chains in experiment is extremely challenging, however, in 
the most recent years, a few techniques to achieve this goal have 
been successfully developed. One 
experimental approach is the formation of chains in so-called mechanically controllable
break junctions. Upon 
pulling two electrodes apart, it is possible to produce short freestanding monatomic 
chains suspended between the electrodes. With this technique, successful chain 
formation has been reported for late $4d$- and $5d$-transition-metal elements and 
transport measurements can be performed to probe the junction properties.
\cite{SmitRuit} Another route is to use a substrate and grow the chains by 
self-assembly exploiting one-dimensional structures provided by the surface topography 
such as step edges\cite{Gambardella} or in trenches of reconstructed surfaces as has 
been demonstrated for Fe chains on Ir(001)\cite{HammerFeIr100} or Au chains on 
Si(111).\cite{Au_chains_on_Si(111)} A second possibility on a surface is to build
the chain atom by atom utilizing a manipulation technique with the tip of a scanning 
tunneling microscope (STM). The use of STM is particularly attractive as it also allows for 
a direct study of the magnetic properties of individual chains. E.g.~in combination with 
spin-polarized STM, Serrate {\it et al.} used this approach to demonstrate non-collinear 
spin alignment in small linear chains of Co atoms on Mn/W(110).\cite{Serrate} On the 
other hand, Hirjibehedin and co-workers created linear Mn chains of up to ten
atoms on an insulating CuN/Cu(001) surface and applied inelastic STM to prove 
their antiferromagnetic exchange coupling.\cite{Mn_CuN_Heinrich}

One of the basic issues in chain magnetism is the sign, origin, and dependence on various 
external parameters of the exchange interaction between the TM atoms. The most 
common way to model a magnetic chain of atoms is the 
effective Heisenberg Hamiltonian:
\begin{equation}
H = - \sum_{n\ne m} J_{nm}\mathbf{S}_n\cdot\mathbf{S}_m,
\end{equation}
which describes the magnetic ground state and low-energy magnetic excitations 
of a system from the knowledge of the 
Heisenberg parameters, or exchange constants, $J_{nm}$. In the latter relation 
$\mathbf{S}_n$ and $\mathbf{S}_m$ are the unit vectors of the spins of atoms 
$n$ and $m$,  i.e.~their magnitude does not depend on their 
relative orientation within this model.
In case of linear equidistant monatomic chains of atoms of the same kind, the 
Heisenberg model can be rewritten with respect to the atom at the origin with the 
Heisenberg constants $J_n\equiv J_{n0}$. The general solution of the Heisenberg 
model on a periodic lattice, here a 1D monatomic chain, is the so-called (flat) 
spin-spiral state, which is a non-collinear arrangement of spins given by:
\begin{equation}
 \mathbf{S}_n=(\cos{(n d q)},\sin{(n d q)},0),
\end{equation}
where $d$ is the spacing between atoms in the chain and $q$ is the modulus of the 
spin-spiral vector. For each set of exchange parameters, the energy dispersion 
$E(q)$ of a spin-spiral with a wave-vector $q$, propagating along the chain axis 
$z$, can be determined from Eq.~(1): 
\begin{equation}
 E(q) = -2\sum_n J_n\cos{(n d q )},
\end{equation}
and the ground state of the system can be found among collinear ferromagnetic (FM, 
$q=0$), collinear antiferromagnetic (AFM, $q=0.5\times 2\pi/d$), and non-collinear 
spin-spiral states. The phase diagram for the 1D-Heisenberg model and characteristic
shapes of the curves $E(q)$ are presented in Fig.~1, where the exchange interaction 
was assumed to vanish beyond third nearest neighbor.

As can be seen from Fig.~1, the phase space of possible ground states of a 1D spin 
chain is rather complicated and depends sensitively on the exchange parameters, 
exhibiting wide regions of preferred non-collinear solutions. As far as real TM chains 
are concerned, establishing their magnetic ground state position in the phase 
diagram of Fig.~1 based on their electronic structure determined by {\it ab initio}
calculations is a non-trivial task. While the exchange interactions have been explored 
intensively for finite TM magnetic clusters, both freestanding and deposited on 
surfaces,\cite{Samir2007,Phivos,Samir2008,Pastor2010,Mn-dimers} the work for 
infinite monatomic TM chains concentrated almost exclusively on collinear magnetic 
solutions.\cite{Y-4dsurface,MokrousovFeIr,MazzarelloFeIr,StepanyukPRB,3d-Tung,
4d-Tung,Urdaniz} The occurrence of non-collinear magnetic states and analysis of 
exchange interaction beyond the first neighbor in freestanding and deposited 
monatomic chains has come into the focus of interest only recently,\cite{Mn-Hafner,
StepanyukJAP,Ataca08PRB} while a conclusive evidence for a spin-spiral ground
state, or more complicated spin textures, in a real deposited metallic TM monatomic 
chain or wire with complex atomic relaxations is missing from {\it ab initio} theory. 
Experimentally, it is very challenging to measure the exchange interaction. 
Nevertheless, the sign of the exchange interaction has been indirectly determined by 
STM for Mn, Fe, and Co dimers on NiAl(110)\cite{PhysRevLett.92.186802} and even 
quantitatively for the interaction between single Co atoms on Pt(111)
\cite{FockoMeier04042008,ZhouNaturePhys2010} and for Mn chains on CuN/Cu(001).
\cite{Mn_CuN_Heinrich} In the two latter cases, indirect exchange by the RKKY 
interaction and super exchange via the substrate atoms, respectively, were concluded 
as the microscopic mechanisms. Surprisingly, for the Mn dimer on NiAl(110) 
ferromagnetic coupling due to double exchange was reported.
\cite{PhysRevLett.92.186802}

  \begin{figure}[t!]
  \begin{center}
    \includegraphics[width=4.2cm]{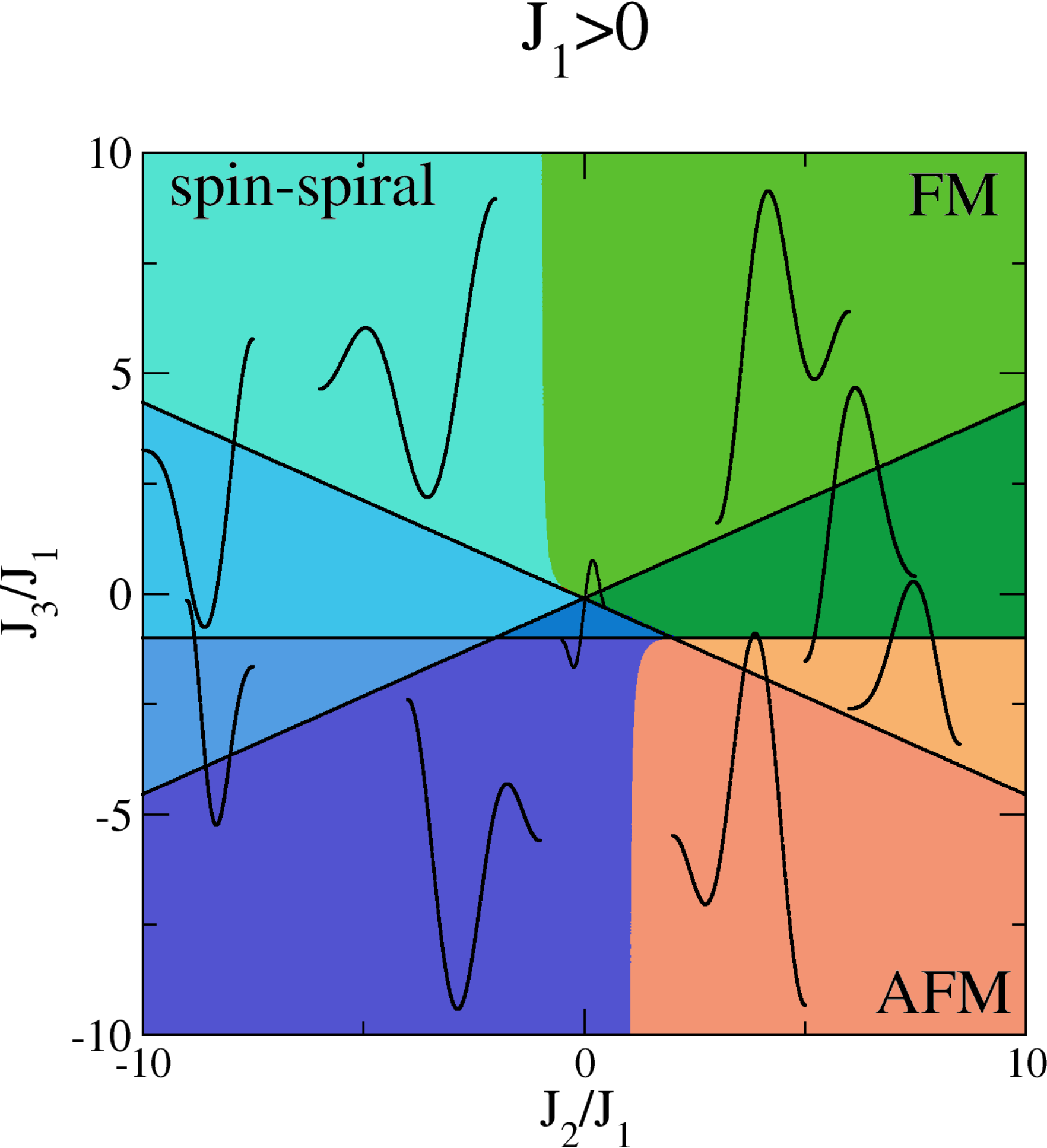}
  \includegraphics[width=4.2cm]{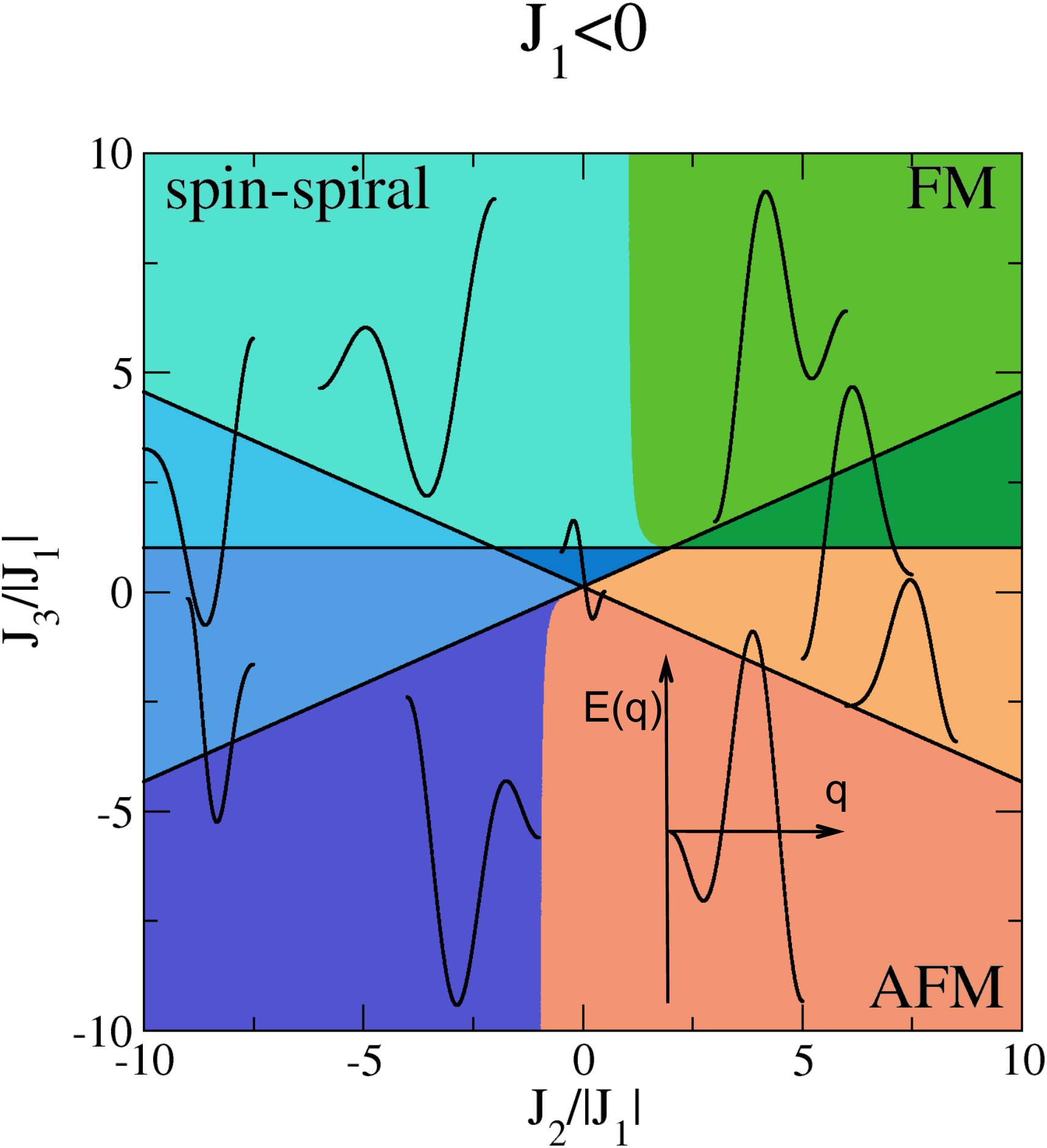}
  \caption{\footnotesize{
    (Color online) 
    Magnetic phase diagram of the 1D Heisenberg model in the $(J_2,J_3)$-parameter 
    space for $J_1>0$ (left graph) and $J_1<0$ (right graph). The cartoons depict the 
    shape of the spin-spiral dispersion curves $E(q)$ in the corresponding phases, with 
    the curve starting at $q=0$ (FM state) and ending at $q=0.5$ (AFM state) in units 
    of 2$\pi$/$d$, where $d$ is the interatomic distance. The $E$ and $q$-axis have 
    been added for one curve for clarity (lower right corner of right graph). In that 
    particular case, the global energy minimum is at the end, i.e.~for $q=0.5$ and 
    the AFM state is most favorable.
  }} \label{phase_diagrams}
  \end{center}
  \end{figure}

Among other $3d$-TMs, Mn chains are most likely to display a manifold of magnetic 
solutions depending on the details of chain geometry, environment and hybridization 
with the substrate. This stems from the fact that a monowire (MW) of Mn atoms 
experiences a FM to AFM ground state transition as a function of the interatomic 
distance $d$ not far away from the equilibrium value $d_0$.\cite{Y-4dsurface} The 
origin of this transition lies in the subtle competition of antiferromagnetic kinetic 
exchange and ferromagnetic double exchange,\cite{AlexanderAndersonModel} which 
in terms of the $d$-electron exchange splitting $\Delta$ and hopping $t$ can be
quantitatively described as a competition between the  $t^2/\Delta$ and $t$ terms, 
respectively.\cite{Phivos} The kinetic exchange arises from a level repulsion
between occupied majority states of an atom with unoccupied minority states of a
neighboring atom, when the spins of the two atoms are opposite. On the other hand,
the hopping between the $d$-states of the neighboring atoms gives rise to an energy 
gain due to the splitting between the bonding and antibonding minority $d$-states, 
if the spins are oriented in parallel, while this splitting is zero in the AFM configuration.

Taking a "prototypical" TM monatomic chain, we normally 
observe smaller values of $\Delta$ and larger values of $t$ for smaller interatomic 
distance $d$, which promotes the AFM kinetic exchange over the FM double 
exchange. Upon increasing the interatomic distance, on the other hand, the hopping 
$t$ is decreased while the exchange splitting energy $\Delta$ increases reaching the 
atomic limit value, which promotes the double exchange over the kinetic exchange, 
and the chain can become ferromagnetic. This results in the Bethe-Slater 
behavior of first exchange parameter $J_1$ for TM chains as a function of 
interatomic distance: $J_1$ is negative for small $d$ and becomes positive for larger 
$d$.\cite{Y-4dsurface} The position of this transition point, where $J_1$ changes 
sign, depends of course on the transition metal, and is situated close to the 
equilibrium interatomic distance only for Mn chains among all 3$d$ TMs.
\cite{Y-4dsurface}

Here, we investigate the appearance of non-collinear magnetism in freestanding 
monatomic Mn chains, its development upon alloying the chains with the 
neighbors of Mn in the periodic table, i.e.~Cr and Fe, and finally upon deposition 
on the Cu(110), Pd(110) and Ag(110) surfaces. Due to the special position of Mn 
chains at the borderline between competing FM and AFM exchange interactions, 
we anticipate the possibility of a spin-spiral ground state in the vicinity of this 
transition point, where $J_1$ is small, and long-ranged indirect exchange 
interactions become of importance.\cite{Ferriani07PRL} Indeed, a whole set of
non-collinear solutions has been recently predicted to occur in single-strand as 
well as thicker infinite and finite freestanding wires of Mn atoms,\cite{Mn-Hafner,
Ataca08PRB} revealing a sensitive dependence of the exchange interactions on 
the details of the surroundings, observed also for small deposited Mn clusters.
\cite{Phivos} Also in two dimensions the tendency of Mn to non-collinear spin 
ordering is well-established.\cite{chiralmagn,ferriani:027201,neel_state,
PhysRevB.72.144420} While experimentally finite Mn chains on CuN are believed 
to display an AFM ordering due to super exchange via substrate atoms, a possibility
for non-collinear spin ordering has been suggested for long finite Mn chains on 
Ni(111) from {\it ab initio} calculations due to frustration of competing exchange 
interactions within the chain and with the surface.\cite{Samir2008} However, 
structural relaxations, which can strongly influence the magnetic order, were not 
considered in that study.

 \begin{figure}
  \begin{center}
  \includegraphics[width=7cm]{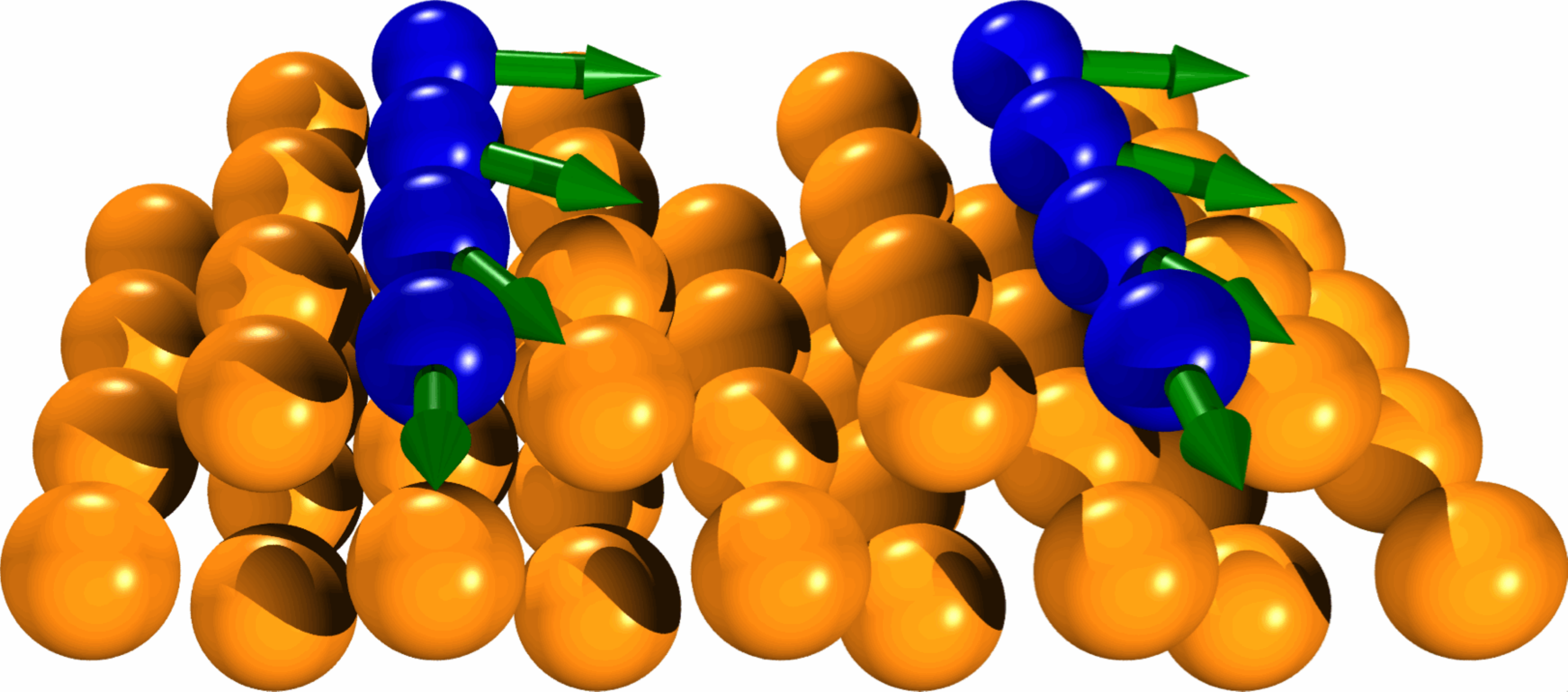}
  \caption{\footnotesize{
    (Color online)
    Geometrical setup for a Mn chain with interatomic distance $d$ deposited on 
    an unreconstructed fcc(110) surface. The blue spheres represent chain atoms 
    with arrows indicating the directions of the spin moments, while the gold 
    spheres represent substrate atoms. The chain atoms are located in the hollow 
    sites of the surface layer above the subsurface layer atoms. The 2D unit cell 
    has $p(2\times 1)$ geometry. A spin spiral state with a value of  $q=0.07\times 
    2\pi/d$ has been chosen for illustration. Note, that in this picture the spin spiral
    rotates within the film plane, however, as we neglect spin-orbit coupling the 
    rotation plane is arbitrary.
  }}
  \label{yura_mncu_pic}
  \end{center}
  \end{figure}

 \begin{figure*}[ht!]
  \begin{center}
  \includegraphics[width=17.5cm]{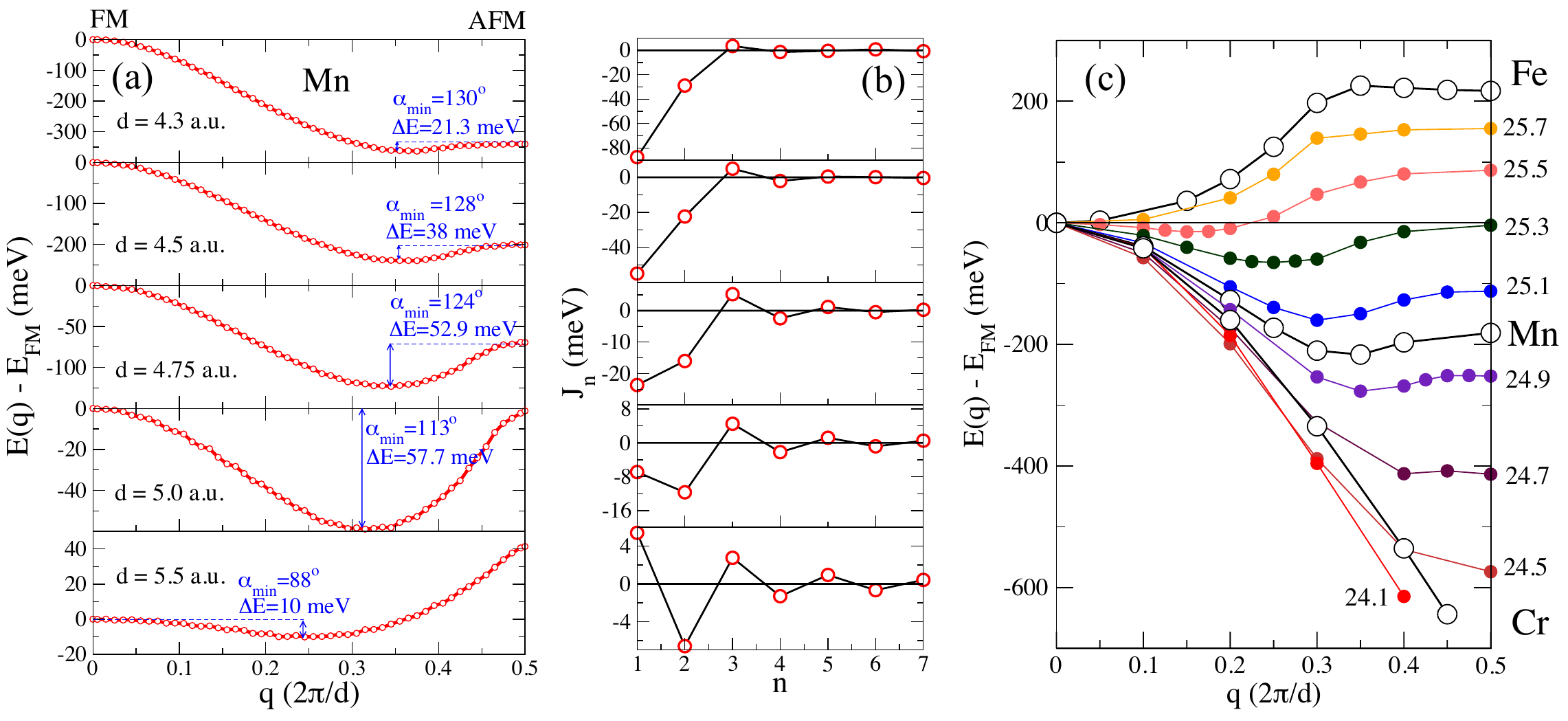}
  \caption{\footnotesize{
    (Color online) 
    (a): Flat spin-spiral dispersions $E(q)$ for freestanding monatomic Mn chains 
    as a function of interatomic distance $d$, calculated using the force theorem.
    The angle $\alpha_{\rm min}$ between spins of nearest neighbor Mn atoms at the
    spiral minimum is given as well as the depth $\Delta E$ of the energy minimum.
    (b): Corresponding Heisenberg exchange constants $J_n$. (c): Flat spin-spiral 
    dispersion $E(q)$ for freestanding chains with an interatomic distance $d=
    4.5$~a.u. and atomic number varying between Cr ($Z=24.0$) and Fe ($Z=26.0$)
    calculated self-consistently. 
    In (a) and (c) the energies are given relative to the ferromagnetic state.
  }} \label{Mn_disp90_pic}
  \end{center}
  \end{figure*}

In our work, we demonstrate that indeed the spin-spiral solution is the ground 
state of freestanding Mn chains in a wide region of interatomic distance $d$ 
between 4.3 and 5.5~a.u. Furthermore, we consider freestanding Mn chains alloyed 
with Cr and Fe atoms and find that such chains take a smooth trajectory in the 
magnetic phase space as a function of Cr, Mn, and Fe concentration. Starting 
from an AFM state for Cr chains the region of non-collinear solutions is crossed 
for Mn-rich chains until with increasing Fe content the FM part of the phase 
diagram is reached. This suggests a simple way to tune the magnetic ground 
state of Mn-containing chains and wires. One can explain this transition within 
the Bethe-Slater\cite{Y-4dsurface} as well as  Moriya picture of direct exchange 
between $d$-states.\cite{Moriya1964} We also investigate whether the 
non-collinear ground state survives for Mn chains on a substrate and consider
the (110)-surfaces of Cu, Pd, and Ag including structural relaxations. Our choice 
of substrates allows us to study the effect of varying interatomic distance in the 
chains as well as the dependence on the strength of hybridization with the 
substrate. We show that owing to the spin-dependent charge transfer between 
the chain and the surface and reduction of the exchange splitting $\Delta$ the 
exchange interaction changes in favor of antiferromagnetic kinetic exchange,
and for Mn chains on Pd(110) the ground state is AFM. On the other hand, a 
shallow spin-spiral ground state in the vicinity of the AFM solution is retained 
for Mn chains on Ag(110) and Cu(110). Since these surfaces are suitable for 
chain creation by self-assembly or by manipulation with an STM tip, we simulate 
STM experiments based on the calculated vacuum charge and magnetization 
density of states. We show that the spin spiral magnetic order in supported Mn 
monowires can theoretically be observed in a spin-polarized STM measurement 
allowing an unambiguous proof of the non-collinear spin structure.

\section{Computational details}

In our first-principles investigation of non-collinear magnetism in monatomic 
Mn chains based on density-functional theory (DFT) we neglect 
spin-orbit coupling, which affects the electronic structure, spin, and orbital 
moments rather weakly for $3d$ TM chains in general, and Mn chains in 
particular, due to their half-filled $d$-shell.~\cite{3d-Tung} Since we
also focus on the metal substrates Cu, Ag, and Pd with relatively small spin-orbit 
coupling this remains a good approximation throughout our study. Neglecting 
the spin-orbit interaction allows for an enormous speed-up of the calculations 
as we need to consider only one TM-atom per unit cell due to the validity of the 
generalized Bloch theorem.\cite{Herring,San86} For all calculations we used the 
GGA rev-\texttt{PBE} exchange-correlation functional.\cite{PhysRevLett.80.890} 
Calculations for the freestanding MWs have been performed using the 
one-dimensional FLAPW method as implemented in the J\"ulich DFT code 
\texttt{FLEUR}.\cite{MokrousovMethod} We used between 64 and 320 
$k$-points carefully checking the convergence of the obtained values with respect 
to their number. The calculation of the spin-spiral energy dispersions was done 
with the magnetic force theorem, and we tested its result versus the self-consistent 
calculations, finding generally very good agreement. For the basis functions cutoff 
we used the $k_{\text{max}}$ parameter of more than 4.0~$\text{a.u.}^{-1}$,
while we chose values of 6~a.u.~and 7.5~a.u.~for the vacuum parameters 
$D_{\rm vac}$ and $\tilde{D}$, respectively. We used a muffin-tin radius of 
$2.1\, \text{a.u.}$ for Mn atoms.

Calculations of the surface-deposited monatomic Mn chains were performed 
with the film version of the \texttt{FLEUR} code~\cite{FLEUR,PhysRevB.69.2005} 
with the geometrical setup and computational details close to those in 
Ref.~[\onlinecite{Y-4dsurface}]. We modeled the semi-infinite crystal by a 
slab of seven substrate layers for the fcc-(110) surfaces of Cu, Pd and Ag. We 
used the experimental lattice constants of the substrates and exploited inversion 
symmetry by depositing chains on both sides of the slab. We restricted structural 
relaxations to the ferromagnetic case. Following values for the muffin-tin radii 
were used: 2.2~a.u.~for Cu atoms, 2.3~a.u.~for Pd and Ag atoms, 2.2~a.u.~for 
the chain Mn atoms. We considered an in-plane separation between the chains of 
approximately 15~a.u.~which corresponds to a $p(2\times 1)$ unit cell (see 
Fig.~2 for a sketch of the geometrical setup). This choice of the super-cell provides 
the separation between adjacent chains large enough to exclude 
the effect of interchain interaction on the spin-spiral energies and energy scales,
discussed in the paper. We chose values of $k_{\rm{max}}
=3.6$ to $3.8~\rm{a.u.}^{-1}$ (depending on the surface) for relaxations and 
non-collinear calculations, achieving the convergence in the values of the total 
energy differences and spin moments. Further computational details, values of 
the spin moments and relaxed atomic positions can be found in 
Ref.~[\onlinecite{Y-4dsurface}]. For the spin-spiral calculations, we considered 
a slab of five substrate layers to simulate the semi-infinite crystal and a dense 
mesh of 336 $k$-points in one half of the Brillouin zone. The calculations of the 
spin-spiral dispersion curves for deposited Mn chains were performed 
self-consistently.

\section{Non-collinear magnetism in free-standing $\text{Mn}$ chains}

It was predicted from first-principles calculations\cite{Y-4dsurface} that 
freestanding Mn MWs exhibit a ferromagnetic to antiferromagnetic ground 
state transition as a function of interatomic distance $d$, a phenomenon 
that has been later predicted to occur also in Mn dimers.
\cite{Mn-dimers} Experimentally, ferromagnetic coupling in Mn dimers on 
NiAl(110) was deduced indirectly from tunneling spectra in STM 
experiments.\cite{PhysRevLett.92.186802} According to the Heisenberg 
model, in the vicinity of the transition point from FM to AFM coupling, the 
nearest-neighbor Heisenberg exchange parameter, $J_1$, becomes 
comparatively small, and exchange constants beyond nearest neighbors
become of importance. This can result in a pronounced non-collinear 
ground state of Mn chains (cf.~phase diagrams in Fig.~1). Indeed, the 
appearance of a spin-spiral energy minimum has been recently 
demonstrated by {\it ab initio} calculations for free-standing chains as 
well as more complex structures of Mn atoms.\cite{Mn-Hafner,
Ataca08PRB} In our work, we investigate how the non-collinear 
magnetism in Mn chains evolves upon stretching the chain and alloying 
with strongly antiferromagnetic Cr and ferromagnetic Fe,\cite{Y-4dsurface} 
alloys that are known to exist in bulk $\alpha$-Mn, 
see~e.g.~Ref.~[\onlinecite{Mn-Williams}].

We start by studying the magnetism in pure Mn chains in the regime of 
the interatomic distance $d$ around the FM-AFM crossover point between 
4.3~a.u.~and 5.5~a.u.\cite{Y-4dsurface} When the spacing is varied in 
this interval, the spin moment of the Mn atoms changes by roughly 
0.3$\mu_B$ with respect to the averaged value of 4$\mu_B$ both for FM 
and AFM solutions. For each value of $d$ the variation of the Mn spin 
moment as a function of the spin-spiral vector $q$ is even smaller (up to 
3\%), which allows a mapping to the Heisenberg model. In 
Fig.~\ref{Mn_disp90_pic}(a) we present the energy dispersion relation 
$E(q)$ of spin spirals (cf.~Eq.(3)) for Mn chains with different interatomic 
spacings $d$. The energies were evaluated with respect to the FM solution,
$q=0$ ($\Gamma$-point), while $q=0.5\times 2\pi / d$ ($X$-point) 
corresponds to the AFM state.

  \begin{figure}[t!]
  \begin{center}
  \includegraphics[width=8.0cm]{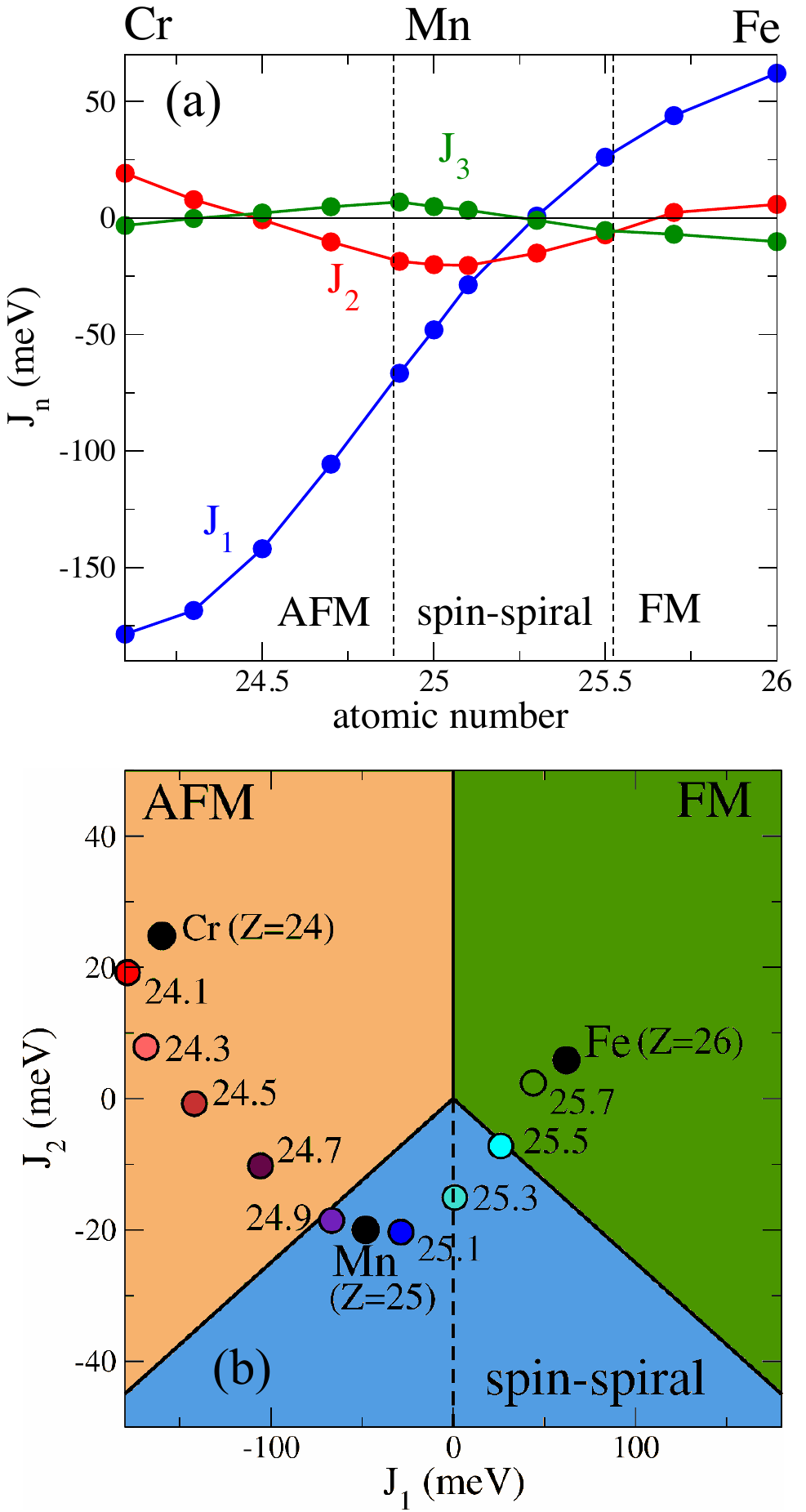}
  \caption{\footnotesize{
    (Color online) 
    (a) Heisenberg exchange constants $J_1$, $J_2$ and $J_3$ as a 
    function of the atomic number $Z$ for freestanding monowires with 
    an interatomic distance of $d=4.5\, \text{a.u.}$ extracted from 
    $E(q)$ given in Fig.~3(c). Dashed lines mark the boundaries of FM, 
    AFM and spin-spiral ground states. (b) Phase diagram of the 1D 
    Heisenberg model in the $(J_1,J_2)$-parameter space. The AFM 
    phase is shaded in beige, the FM phase in green and the spin-spiral 
    phase is colored in blue. The circles represent the values of the 
    Heisenberg exchange constants $J_1$ and $J_2$ given in (a).
  }} \label{frac_phase_diag_pic}
  \end{center}
  \end{figure}

Starting already at $d=4.3$~a.u.~a very shallow spin-spiral energy minimum 
of $\Delta E=21.3$~meV with respect to the AFM state develops at $q\approx 
0.35\times 2\pi/d$, corresponding to an angle of approximately $130^\circ$ 
between spins of nearest neighbor Mn atoms. 
As the interatomic distance increases this energy 
minimum becomes deeper, reaching as much as 57.7~meV for a distance of 
5.0~a.u., and moves towards the middle of the BZ with the angle between 
the spins on neighboring atoms, $\alpha_{\rm min}$, becoming smaller. At 
this distance, the difference in energy between the FM and AFM state, 
negative for smaller $d$, nearly vanishes, and upon further increasing $d$, 
when the FM state moves to lower energies with respect to the AFM solution, 
the non-collinear ground state energy minimum becomes less pronounced 
again. Remarkably, in the whole range of considered interatomic distance 
the ground state of the Mn chains is non-collinear.

In Fig.~3(b) we display the Heisenberg exchange constants $J_n$ as a
function of the $n$th nearest neighbor in the chain extracted from the 
corresponding $E(q)$ dispersion curves in Fig.~3(a) by a Fourier transform.
At a lattice 
constant of 4.3~a.u.~$J_1$ is large and negative, indicating AFM exchange 
coupling that dominates over all other Heisenberg constants, which can be 
considered negligible beyond the 2nd nearest neighbor. Among the collinear 
solutions the AFM state is therefore favored over the FM state by a large 
value of 340~meV and the spin-spiral minimum is shallow. With increasing 
$d$ the long-range exchange interaction becomes more significant as $J_1$
decreases drastically until it changes its sign to ferromagnetic coupling 
between 5.0 and 5.5~a.u. Between these values of $d$, the 2nd and 3rd 
nearest exchange constants, $J_2$ and $J_3$, become of importance and 
$J_2$ even exceeds the value of $J_1$, resulting in a deep non-collinear 
ground state. Finally, at $d=5.5$~a.u.~the nearest-neighbor exchange 
interaction is ferromagnetic and in the long-range interaction the typical 
RKKY-like behavior is clearly visible. At this and larger distances $d$ the 
magnitude of the exchange constants, the depth of the spin-spiral ground 
state, and the energy difference between FM and AFM state becomes small 
due to a large separation between the atoms.

The transition from AFM to FM nearest-neighbor exchange coupling with 
increasing interatomic spacing is predicted from the Bethe-Slater curve 
(see e.g.~Ref.~[\onlinecite{SimpleModelsMagnetism}]). The mechanism 
behind this effect can be understood within the Alexander-Anderson model
\cite{AlexanderAndersonModel} of exchange interaction between the 
$d$-states of transition-metal atoms as applied by Moriya.
\cite{Moriya1964} At small distances, the kinetic exchange interaction, 
which is antiferromagnetic, dominates for a TM atom with nearly half-filled 
$d$-shell such as Mn. As the spacing is increased, the hopping $t$ and 
splitting between bonding and antibonding parts of the $d$-states decrease,
while exchange splitting $\Delta$ normally increases. This leads to weakening
of the kinetic exchange according to $t^2/\Delta$ dependence, outlined in
the introduction. On the other hand, the ferromagnetic double exchange,
whose energy contribution is proportional to $t$, is also weakened upon 
stretching, but only linearly in the hopping. This leads to a crossover between 
negative $t^2/\Delta$ and positive $t$ parts of the exchange energy
at a certain interatomic distance, and the chain becomes ferromagnetic
upon further stretching.  
 We will see in the next 
section how this competition is changed in favor of kinetic exchange due 
to hybridization with a substrate.

Within the so-called virtual crystal approximation (VCA) we artificially 
change the atomic number $Z$ of the atoms in the chain while keeping the 
charge neutrality, and investigate the influence of the $3d$-band filling on 
the non-collinear magnetism in Mn MWs. By doing this we aim at mimicking 
the behavior of the magnetic ground state upon alloying Mn with other 
elements, as well as charging upon~e.g.~deposition on a substrate. We 
perform these calculations 
at a fixed interatomic distance in the chain of $d=4.5$~a.u., roughly 
corresponding to the equilibrium interatomic distance of the TM chains in 
the middle of the $3d$ series.\cite{3d-Tung} The atomic number $Z$ is 
varied between 24.0 and 26.0, corresponding to Cr and Fe chains, 
respectively, in steps of $\Delta Z=0.1-0.2$, which causes a shift of the 
Fermi level $E_F$ within the $3d$-bands upwards in energy with respect 
to the Fermi energy in the Cr chain.

The results of our calculations for the spin-spiral dispersion $E(q)$ are 
presented in Fig.~3(c) as a function of $Z$. We observe, that starting 
off with Cr ($Z=24$), which has a pronounced AFM ground state with 
an energy gain of more than 0.6~eV with respect to the FM state, the 
energy difference between the FM and AFM state drops rapidly 
upon increasing the atomic number and vanishes at $Z\approx 25.3$.
Upon further increasing $Z$, the FM state becomes favorable over the 
AFM configuration, and for an Fe chain with $Z=26$, the FM ground 
state is by more than 0.2~eV~lower in energy than the AFM solution, 
in accordance to previous collinear calculations at this distance.
\cite{Y-4dsurface} At values of $Z$ around 24.9 a small non-collinear 
ground state emerges close to the $X$-point, which shifts towards the 
$\Gamma$-point and becomes more pronounced in energy with 
increasing $Z$ until it disappears at $Z\approx 25.6$. The energy gain 
of the spin-spiral ground state with respect to the collinear solutions 
reaches as much as 75~meV for $Z\approx 25.3$, compared to 
38~meV~in a Mn MW with $Z=25$ (c.f.~Fig.~3(a)). Overall, we 
observe a smooth transition between the dispersions of Cr and Fe 
chains.

  \begin{figure}[t!]
  \begin{center}
  \includegraphics[width=8.7cm]{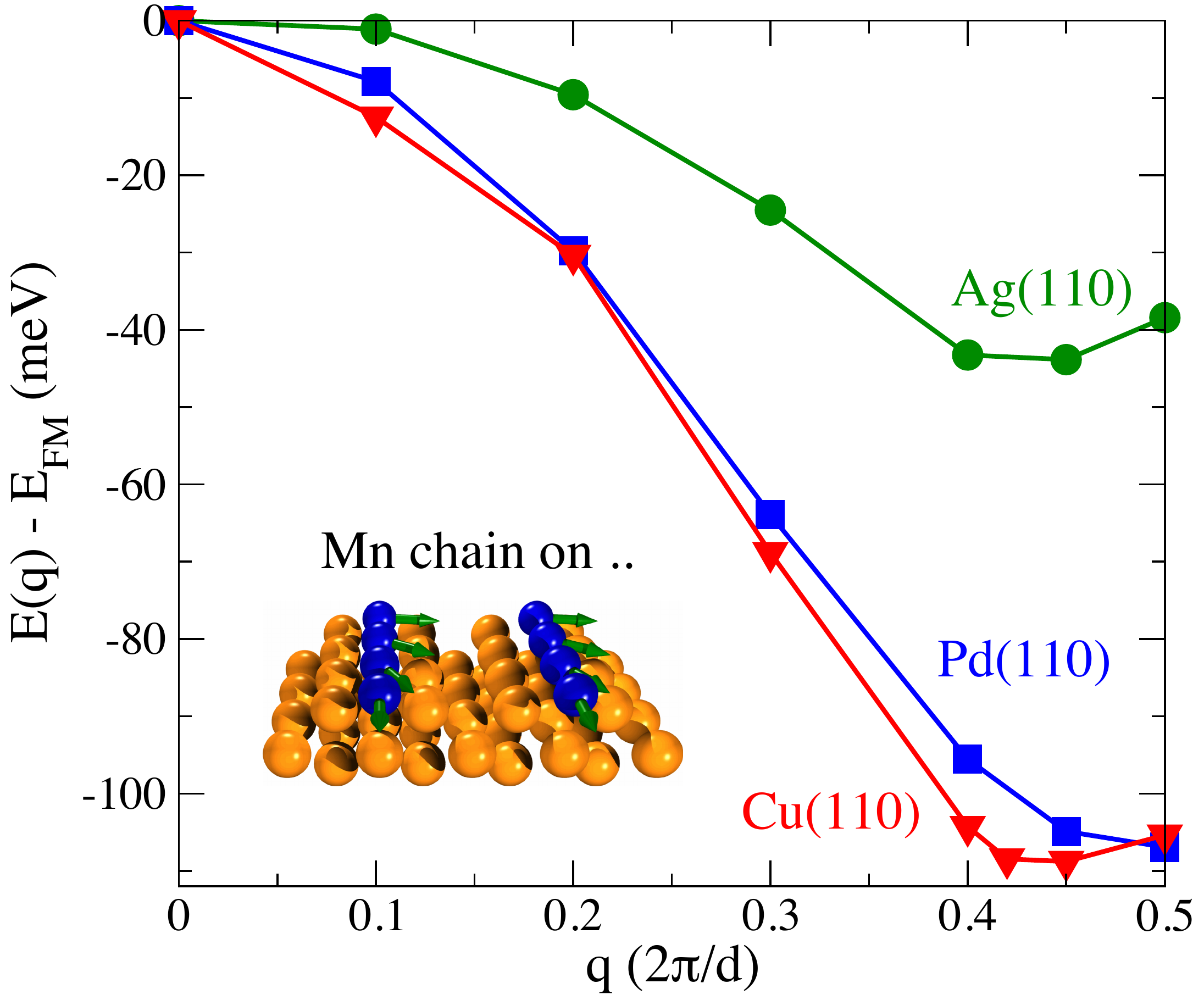}
  \caption{\footnotesize{
    (Color online) 
    Self-consistently calculated flat spin-spiral dispersion curves $E(q)$ 
    for a Mn chain on Cu(110) (red triangles), Ag(110) (green circles) 
    and Pd(110) (blue squares).
  }} \label{dispersion_all_sub_pic}
  \end{center}
  \end{figure}

From the spin-spiral dispersions we extract the Heisenberg exchange 
parameters for the first three neighbors, $J_1$, $J_2$, and $J_3$ as 
a function of $Z$, presented in Fig.~4(a), and analyze the trajectory 
a monatomic chain takes in the magnetic phase space when its 
atomic number is changed between 24 and 26 as shown in Fig.~4(b). 
From this plot we can see that mostly $J_1$ exceeds $J_2$ by an 
order of magnitude, while, in turn $J_3$ is noticeably smaller than 
$J_2$ $-$ thus, for our analysis we will consider only $J_1$ and $J_2$ 
Heisenberg parameters and trace the position of our chains in the 
$J_1$-$J_2$ phase space. The magnetic ground state for the 1D 
Heisenberg model as a function of $J_1$ and $J_2$ is presented 
in Fig.~4(b). 

We observe in Fig.~4(b) a smooth trajectory of the MW in the 
$J_1$-$J_2$ phase space. In the regime of $Z$ close to the pure Cr 
chain antiferromagnetic nearest-neighbor exchange coupling prevails, 
i.e.~$J_1$ is about $-150$~meV and very large in comparison to 
$J_2$ with an absolute value of approximately 20~meV. 
Accordingly, in this region of $Z$ the ground state is AFM and the 
MWs are situated deeply inside the AFM region in the phase diagram
of Fig.~4(b). When $Z$ approaches a value of 24.9, $J_1$ drops 
significantly while $J_2$ is strengthened so that both exchange 
parameters are becoming closer in value. In the phase space, the 
chain therefore approaches the boundary between the AFM and 
spin-spiral phases. This boundary is crossed when $Z$ exceeds 24.9 
and the chains reveal a non-collinear ground state until $Z$ reaches
25.5. In terms of exchange constants this happens when upon 
increasing atomic number, $J_1$ is dramatically decreased so that it 
becomes comparable to $J_2$. At $Z\approx 25.3$ $J_1$ vanishes,
then it becomes positive and increases, exceeding $|J_2|$ for 
$Z>25.5$ $-$ correspondingly, the chain becomes ferromagnetic for 
$Z$ in the vicinity of Fe. 

Remarkably, the dependence of $J_1$ on $Z$, displayed in Fig.~4(a), 
is very similar to that of the Bethe-Slater curve, which characterizes 
the sign of the exchange interaction as a function of $d/r_d$, with $d$ 
being the distance between the transition-metal atoms and $r_d$ as 
spread of the $d$-orbitals. It was demonstrated from {\it ab initio} 
calculations in Ref.~[\onlinecite{Y-4dsurface}] that the $3d$ 
transition-metal monatomic chains display the Bethe-Slater behavior 
as a function of $d$, with chain's position on this curve corresponding 
to $r_d$,~c.f.~Fig.~3 in Ref.~[\onlinecite{Y-4dsurface}]. The 
Bethe-Slater-like behavior of $J_1(Z)$ in Fig.~4(a) is another evidence 
of the applicability of this simple picture for $3d$ TM chains: at fixed 
interatomic distance $d$ (of 4.5~a.u.) upon varying the nuclear charge 
$Z$ the spread of $d$-orbitals is changed. For smaller $Z$ the $r_d$ 
is larger and for larger $Z$ the spread is smaller\cite{Slater:1930} 
$-$ this results in positioning of the AFM Cr and FM Fe chains on different 
sides of the Bethe-Slater curve, with Mn chains in the middle at the 
FM-AFM crossover.

A different point of view can be taken by referring to the model of 
Moriya,\cite{Moriya1964}, which is an extension of the Alexander and
Anderson model~\cite{AlexanderAndersonModel} for the exchange 
interaction between transition metals as a function of the number of 
electrons in the $d$-shell. Moriya finds a curve similar to that of 
Bethe-Slater if he varies the number of $3d$-electrons of the 
interacting transition metal atoms, in complete analogy to $J_1(Z)$ 
presented in Fig.~4(a). This confirms that while the kinetic exchange
mechanism is behind the AFM ordering of TMs with half-filled $d$-shell, 
the double exchange leads to FM ground state for TMs with half-filled 
$d$-spin-subband. The link between the Bethe-Slater and Moriya 
pictures of exchange, as outlined in the previous paragraph, is the 
decrease of the spread of the $d$-orbitals, $r_d$, upon increasing 
electronic occupation in TM series.\cite{Slater:1930} Effectively, as 
far as Mn chains are concerned, we can therefore conclude that the 
effect of the band filling on the magnetic state is the same as that 
from changing the interatomic distance $d$ in the chain.

For the next-nearest and next-next-nearest neighbor exchange 
interactions, $J_2$ and $J_3$, we find two changes of sign in the 
considered range of atomic numbers,~c.f.~Fig.~4(a). This result is in nice qualitative
agreement with the spin susceptibilities for bulk Fe calculated by 
Terakura et al.~\cite{Terakura1982} as a function of the energy. 
As the increase of the atomic number in our approach leads to a filling
of the $3d$-band we can compare our result with the spin 
susceptibilities.

\section{Non-collinear magnetism in deposited $\text{Mn}$ chains}

For supported Mn chains we consider three different metallic surfaces: 
unreconstructed Cu(110), Ag(110) and Pd(110). This choice is 
motivated by several reasons. Firstly, these substrates are rather 
promising experimentally as they provide trenches for chain self-assembly, 
while
chain creation atom-by-atom by 
manipulation with an STM tip may also be feasible. Secondly, as far as
we assume that the chains grow pseudomorphically on these surfaces, 
the interatomic distance between the chain atoms $d$ is dictated by 
the lattice constant of the substrate, taking values of 4.82, 5.30 and 
5.59~a.u.~for Cu(110), Pd(110) and Ag(110), respectively. In the 
previous section we observed that in this regime of interatomic 
distance Mn chains display a non-collinear ground state, while among 
the collinear solutions freestanding Mn chains favor an AFM  
state for $d$ corresponding to that of Cu(110), while for the spacing 
corresponding to Pd(110) and Ag(110) the chains prefer a ferromagnetic 
configuration. Finally, while Cu and Ag are noble metals with 
predominantly $s$-character of electronic states at the Fermi level, the 
degree of hybridization of Mn atoms and the underlying Cu or Ag 
substrate is much weaker than that of Mn chains deposited on Pd, 
which has high density of states of $4d$ electrons at the Fermi energy. 
We can thus compare the effect of hybridization on the exchange 
interaction of deposited Mn MWs.

  \begin{figure}[t!]
  \begin{center}
  \includegraphics[width=7.9cm]{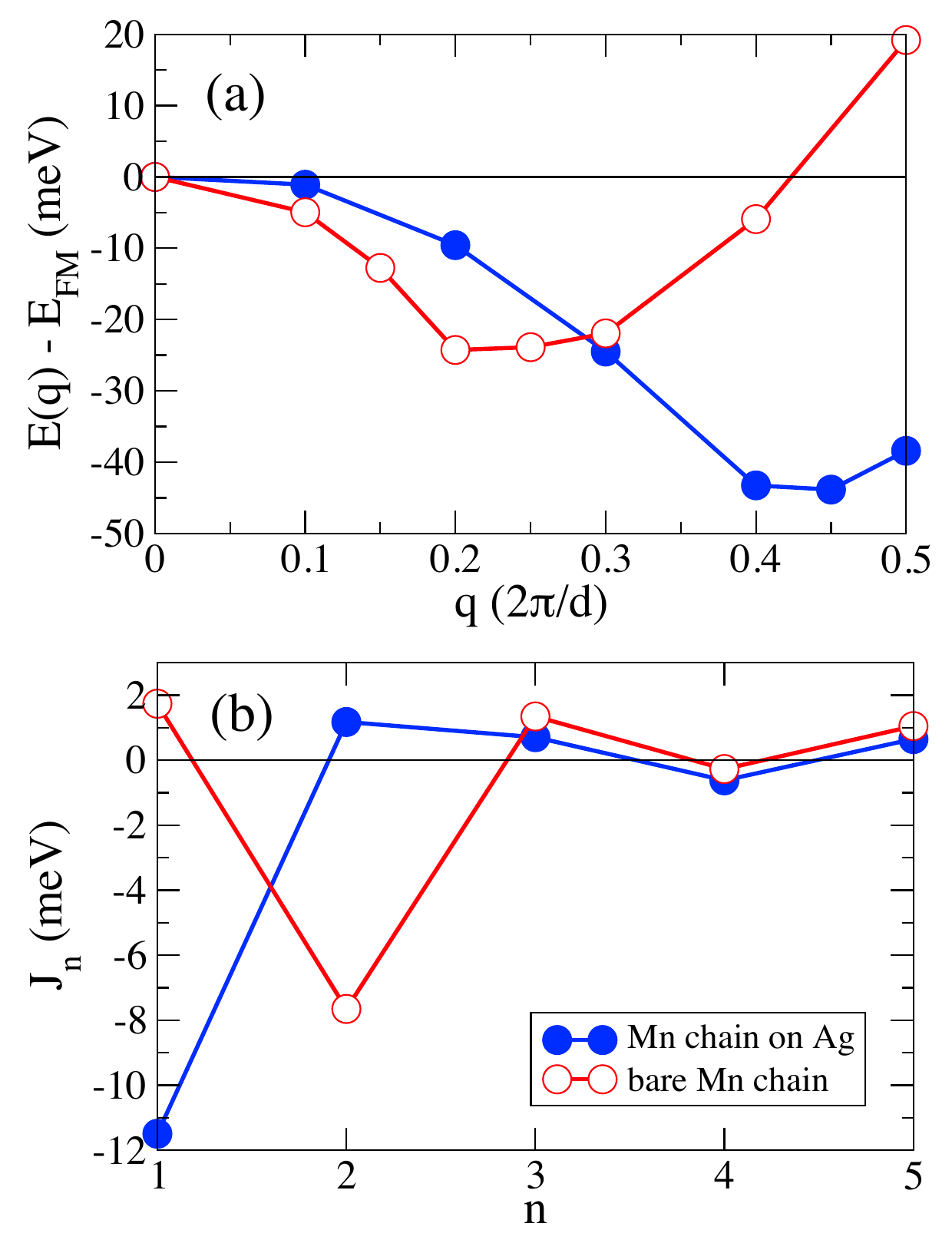}
  \caption{\footnotesize{
    (Color online) 
    (a) Flat spin-spiral energy dispersion calculated self-consistently for 
    a Mn chain on Ag(110) (full blue circles) and a freestanding Mn 
    monowire with the same interatomic spacing of $d=5.59 \, \text{a.u.}$ 
    (empty red circles). (b) Heisenberg exchange constants extracted from 
    spin-spiral dispersion curves in (a). Full blue circles show the
    $J$'s for a Mn chain on a Ag(110) surface and open red circles represent 
    the Heisenberg constants for a freestanding Mn wire with the same 
    interatomic distance of $d=5.59 \, \text{a.u.}$
  }} \label{MnCu_dispersion_pic}
    \end{center}
  \end{figure}

The summary of our calculations is presented in Fig.~5, where the 
energy dispersion curves $E(q)$ are plotted for Mn chains on Cu(110), 
Pd(110) and Ag(110), as a function of the $q$-vector of a flat 
spin-spiral propagating along the chain. We observe that in all cases 
the energetically favorable state is AFM among the collinear solutions, 
even for Mn chains on Pd(110) and Ag(110), which prefer FM ordering 
in an unsupported situation. This tendency towards antiferromagnetism in 
surface-deposited chains and small clusters of $3d$ TMs has been observed 
previously in Refs.~[\onlinecite{Phivos}] and [\onlinecite{Y-4dsurface}],  
and stems from the hybridization between the chain and the substrate. 
As far as the non-collinear solutions are concerned, we predict that 
despite strong modifications in the shape of $E(q)$ and a tendency 
towards AFM-ordering, a shallow spin-spiral energy minimum retains 
for Mn chain on Cu(110) (3.5~meV below the AFM state) and on 
Ag(110) (5.5~meV below the AFM solution). In contrast, the strong 
hybridization between the $4d$-states of the Pd surface and the 
$3d$-states of Mn atoms around the Fermi energy leads to a 
disappearance of the spin-spiral minimum, and the resulting 
order is antiferromagnetic.

In order to investigate the influence of the substrate on exchange 
interactions of deposited Mn chains in more detail, we choose the most 
striking example of a Mn MW on Ag(110). In Fig.~6(a) we present the 
self-consistently calculated flat spin-spiral dispersion curve for a Mn 
chain on Ag(110) in comparison to the dispersion curve of a 
freestanding Mn chain at the same interatomic distance of 
$d=5.59$~a.u. We observe that while the freestanding Mn monowire 
exhibits a significant energy minimum of $\Delta E= 25\, \text{meV}$ 
with respect to the FM solution of a spin-spiral with an angle of 
$\alpha_{\text{min}}=72^\circ$ between neighboring spins, the 
spin-spiral state becomes much less favorable for the chain on the 
Ag(110) substrate ($\approx 5.5\, \text{meV}$ below the AFM state)
and the angle shifts to a larger value of $\alpha_{\text{min}}=
151^\circ$. As already pointed out, the energy difference between the 
FM and the AFM solution is reversed for a Mn chain deposited on the 
substrate. While the FM state is by 20 meV lower than the AFM for the 
freestanding Mn chain, the AFM becomes by 40 meV more favorable 
for the deposited Mn chain. The comparison of the exchange 
parameters, Fig.~6(b), which were extracted from the dispersion curves 
in Fig.~6(a), underlines the influence of the substrate on the nature of 
exchange interactions in this system: while the exchange constants 
beyond the second nearest-neighbor are similar for the freestanding 
and Ag-supported Mn chain, the Ag(110) surface causes a change in 
sign of the parameters $J_1$ and $J_2$, as compared to the 
freestanding Mn MW. Further, the magnitude of $J_1$ increases 
significantly, emphasizing predominantly AFM-coupling along the
deposited chain.

  \begin{figure}[t!]
  \begin{center}
  \includegraphics[width=8.9cm]{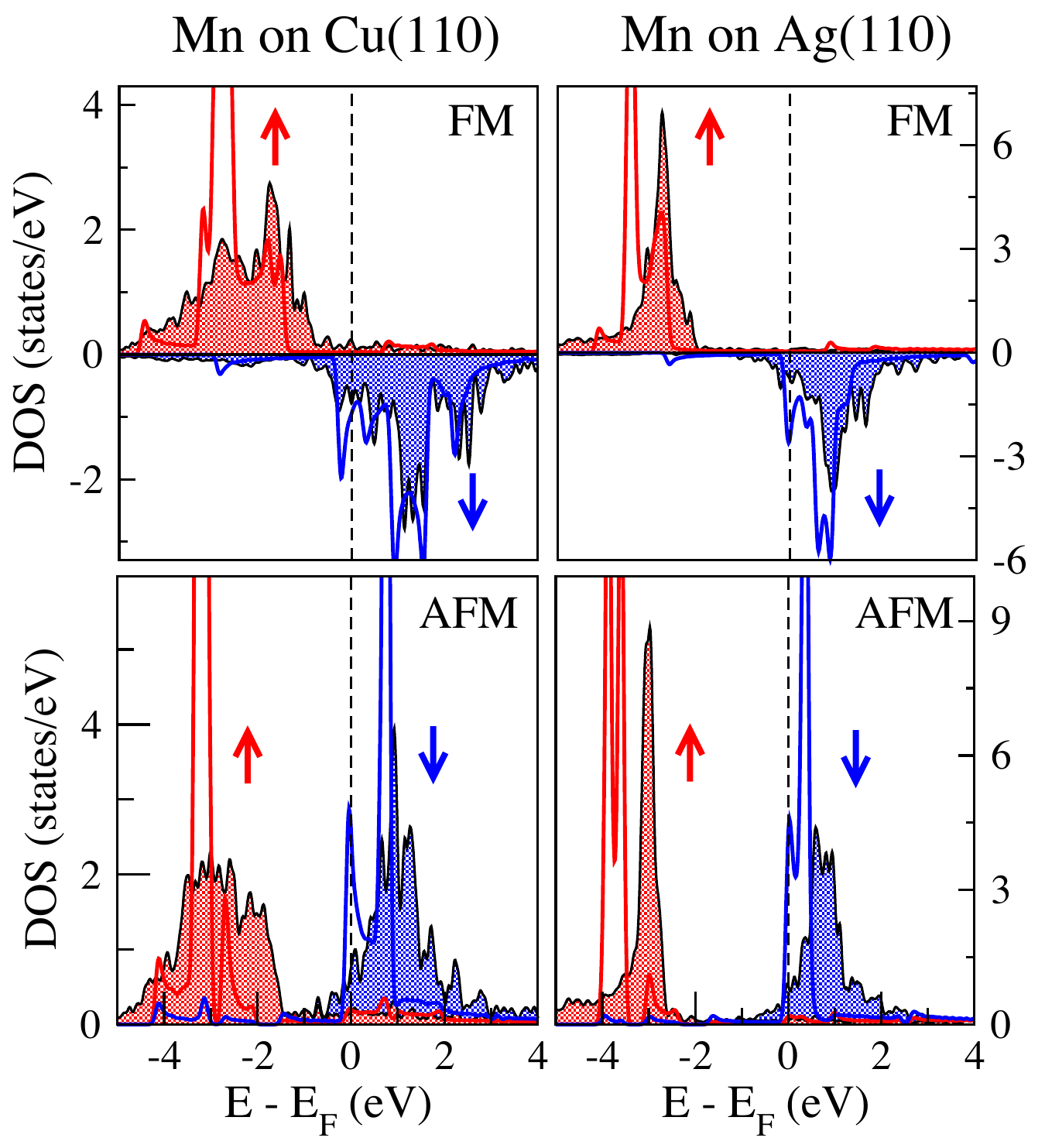}
  \vspace{0.3cm}
  \caption{\footnotesize{
    (Color online) 
    Local density of states (DOS) of a Mn atom in a chain deposited on 
    Cu(110) (left panel) and Ag(110) (right panel) in comparison to the 
    Mn DOS of a bare chain at the corresponding lattice constant, both 
    for ferromagnetic (FM) and antiferromagneic (AFM) arrangement of 
    the spins. While the DOS of the freestanding chains is given with a 
    solid line, the DOS of the deposited Mn chains corresponds to the 
    shaded areas. Red (blue) lines, red (blue) shading and red (blue) 
    arrows mark the spin-up (spin-down) DOS.
  }} \label{Mndos}
    \end{center}
  \end{figure}

It seems fruitful to analyze the correspondence between the changes 
in the local Mn spin moment $\mu_S^{\rm Mn}$ upon deposition and
the change in the nature of exchange coupling. In general, as it was 
already reported in Ref.~[\onlinecite{Y-4dsurface}] for the FM and AFM 
state of Mn chains on the (110)-surfaces of Cu, Ag, and Pd, the 
value of $\mu_S^{\rm Mn}$ (in units of $\mu_B$) is reduced in the 
FM (AFM) state from 3.97 (3.98) for an unsupported Mn chain to 3.69 
(3.75) on Cu(110), from 4.12 (4.20) for an unsupported Mn chain to 
4.00 (3.96) on Pd(110), and from 4.20 (4.29) for an unsupported Mn 
chain to 4.05 (4.12) on Ag(110). This reduction of the spin moment 
comes from spin-dependent changes in the number of electrons inside 
the Mn atomic spheres. While the changes in the local Mn spin moment 
stemming from $s$-electrons are rather small upon deposition, owing to 
small spin-polarization of these states, the changes in the $d$-moment 
and $d$-occupation for up and down spins are rather significant. For 
example, in the case of the AFM Mn chain the number of $d$-electrons
in the spin-up channel is changed from 4.47 in the freestanding state 
to 4.39 upon deposition on Ag(110), while the corresponding values 
for the spin-down channel constitute 0.27 and 0.37, respectively.

  \begin{figure}
  \begin{center}
 \includegraphics[width=8.5cm]{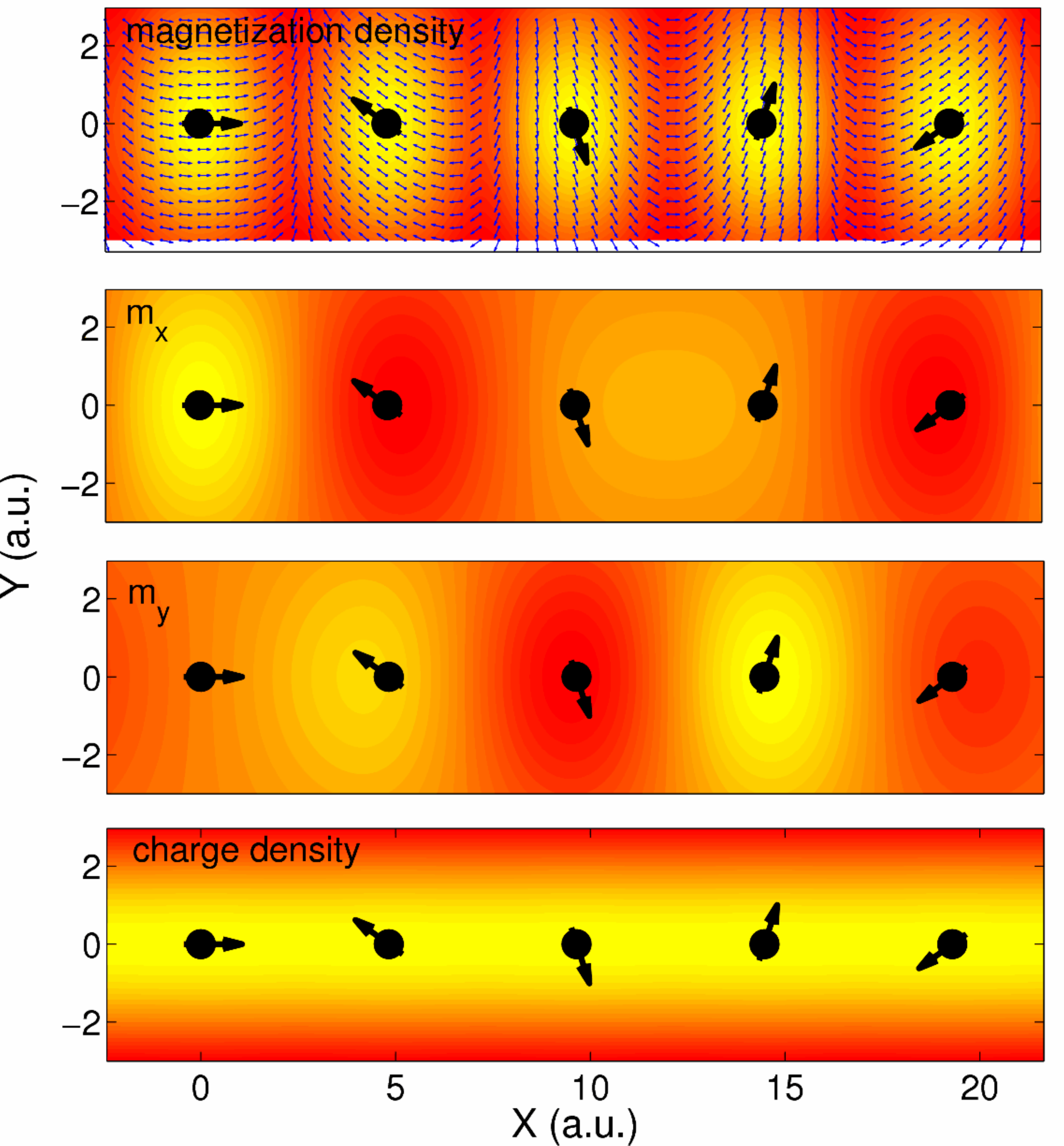}
  \caption{\footnotesize{
    (Color online) 
    Local vacuum charge and magnetization density of a Mn chain 
    on Cu(110) in a spin-spiral state with $q=0.4\times \frac{2\pi}{a}$ 
    at a height of $z=5.9 \, \text{\AA}$ above the chain and for an 
    energy interval of $E=[E_F -300 \, \text{meV}, E_F]$. The chain 
    extends along the $x$-direction and the black circles represent the 
    position of the Mn atoms with arrows indicating their magnetic 
    moments. The top panel shows the magnetization density: whereas 
    the background color denotes the absolute value of the local 
    magnetization (yellow (red) corresponding to high (low) values), the 
    small blue arrows indicate the unit vector of its direction. The second
    and the third panels display the $x$- and $y$-component of the 
    magnetization, respectively, while the bottom panel shows the 
    vacuum charge density.
  }} \label{dens_0.4_pic}
  \end{center}
  \end{figure}

The values for the changes in the Mn $d$-occupation upon deposition
are quite similar for other substrates and magnetic states of deposited 
Mn chains and a general trend emerges: upon deposition we observe a 
decrease in the occupation of spin-up electrons, while the number of 
spin-down electrons increases, with consequent reduction of the total 
Mn spin moment. In a simple picture of the densities of states (DOS) 
of a freestanding chain, this situation corresponds to an effective 
decrease in the value of the interatomic distance $d$, when the chain 
is deposited on a surface: the spin-down $d$-states (mostly unoccupied 
in Mn) are shifted to lower energies while the spin-up $d$-states (mostly 
occupied in Mn) are shifted in the opposite direction, and the overall 
exchange splitting and the value of the spin moment is reduced. 
Qualitatively, as we discovered in the previous section, this reduction in 
interatomic distance has the same effect as the alloying of the Mn wire 
with Cr atoms, which results in large and negative $J_1$ exchange 
parameter, see Figs.~3(b) and 4(b), and the exchange interaction in 
deposited Mn chains leans towards strong antiferromagnetism, as 
observed in Fig.~6.

The reduction of exchange splitting $\Delta$ of the Mn $d$-states upon 
deposition on a substrate can be seen also from the local densities of 
states of deposited chains in comparison to the DOS in the freestanding 
configuration. In Fig.~\ref{Mndos} we compare the DOS of bare and 
Mn chains deposited on Cu(110) and Ag(110) substrates for both 
collinear spin configurations, i.e.~FM and AFM states. The reduction of 
$\Delta$, defined as the difference between the center of mass of spin-up
(red) and spin-down (blue) Mn states, upon deposition on the surface can 
be easily seen in these plots. It is especially pronounced for the case of a 
Mn chain deposited on Ag(110), both in the FM and AFM configuration. 
In this case the distance $d$ between the Mn atoms along the chain is 
larger and the corresponding peaks in the DOS are sharper than in the 
case of the Mn chain deposited on Cu(110) which imposes a smaller 
Mn-Mn distance. As can be seen, upon deposition on Ag(110) the value 
of the exchange splitting can be reduced by as much as 0.3$-$0.4~eV 
as compared to the freestanding value of $\Delta$. In case of Cu(110) 
the Mn states are rather spread in energy for both structural cases and 
the center of mass of the $d$-bands is more difficult to determine, however, 
even in this case the reduction of the exchange energy upon deposition 
is quite visible. 

In terms of the Alexander-Anderson model, outlined in the previous 
sections, the reduction in the value of $\Delta$ promotes the 
antiferromagnetic kinetic exchange over ferromagnetic double exchange
between the Mn atoms. Another subtle point
in this competition, which was pointed out in Ref.~[\onlinecite{Phivos}],
is that the gain in energy due to double exchange is smaller for
a chain on a surface than it is for a free-standing chain. If the $d$-states
are resonant with the background hybridization, e.g.~due to the presense
of a substrate, the tails of the bonding and antibonding resonances 
cross the Fermi energy, and their repopulation partly counteracts the 
energy gain due to double exchange in a bare chain.\cite{Phivos}
At the end, due to enhancement of the kinetic exchange and weakening
of the double exchange, Mn chains, situated right at the FM-AFM crossover 
point, become antiferromagnetic when deposited on a substrate. 

  \begin{figure}
  \begin{center}
 \includegraphics[width=7.8cm]{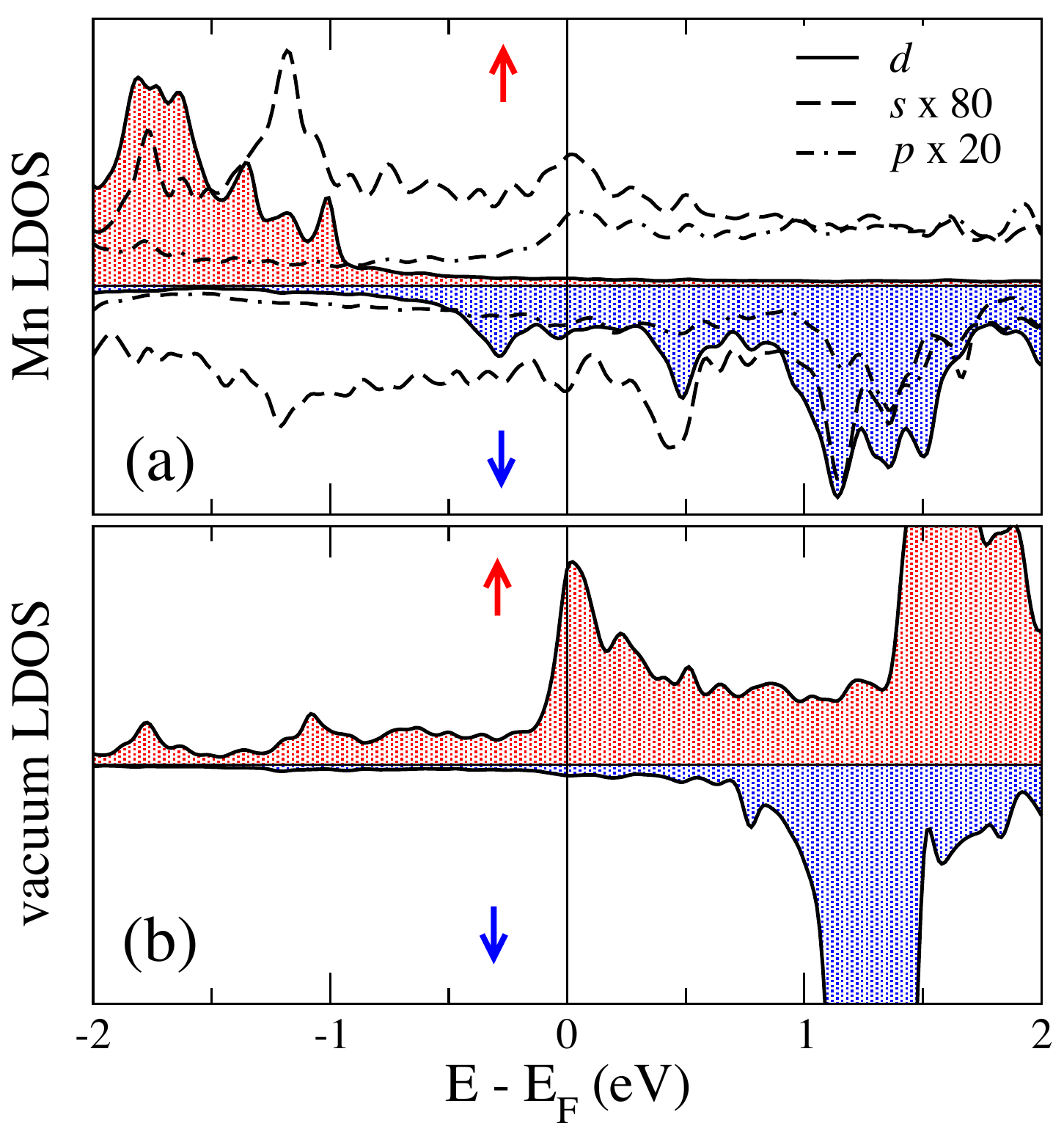}
  \caption{\footnotesize{
    {(Color online) 
    (a) Local density of states of a Mn atom in a ferromagnetic chain 
    deposited on Cu(110) decomposed into $s$-,$p$- and $d$-contributions. 
    The $s$- and $p$-DOS is multiplied by a factor of eighty and twenty, 
    respectively. (b) Corresponding local DOS in the vacuum at a height 
    of $z=5.9$~\AA~above the chain. In both plots, red (blue) shading 
    and red (blue) arrows mark the spin-up (spin-down) DOS. 
    The units are arbitrary.}
  }} \label{vacuum_dos}
  \end{center}
  \end{figure}

\section{Simulation of spin-polarized STM images of non-collinear 
$\text{Mn}$ chains}

An experimental verification of the predicted non-collinear magnetic 
ground state of Mn chains on Ag(110) and on Cu(110) is very 
challenging as the atomic chain structure needs to be revealed and
individual chains have to be addressed. In addition, the magnetic 
signal is small due to the low amount of magnetic material and in 
any spin-spiral state the magnetization is compensated on the
atomic length-scale. Therefore, there are few experimental 
techniques suitable for this challenge. In recent years, it has been 
demonstrated that spin-polarized scanning tunneling microscopy 
(SP-STM) is a powerful tool to image magnetic order of individual 
nanostructures down to the atomic-scale~\cite{kubetzka:087204,
Yayon07PRL,Limot08PRL,Serrate} and non-collinear spin structures
in ultra-thin films have been observed.~\cite{chiralmagn,
ferriani:027201,GWK2008,Wasniowska10PRB} As calculated SP-STM 
images based on the spin-polarized extension of the Tersoff-Hamann 
model~\cite{D_Wortmann} are in excellent agreement with the 
experimental data, we have performed such simulations for the 
system of deposited atomic chains, taking Mn chains on Cu(110) 
as an example.

Before we turn to the predicted SP-STM images, it is insightful to 
analyze the partial charge and magnetization density obtained from 
our first-principles calculations in the vicinity of the Fermi energy. 
As we are interested in the contrast obtainable with SP-STM, we 
depict these quantities in the vacuum a few {\AA}ngstr\"om above 
the Mn chain atoms. Figure~\ref{dens_0.4_pic} displays such a top 
view of the Mn chain on Cu(110) with a spin-spiral magnetic state 
characterized by the vector $q=0.4\times \frac{2\pi}{a}$. This 
magnetic structure is close to the absolute minimum in total energy
(see Fig.~\ref{dispersion_all_sub_pic}), and it is commensurate 
with the atomic lattice with a relatively small pitch of 5 atoms 
allowing a simple discussion.

In the magnetization density (top panel of Fig.~\ref{dens_0.4_pic}) 
the location of the Mn atoms can be clearly seen while the vacuum 
charge density (bottom panel) only resolves the chain direction. The 
$x$- and $y$-direction of the magnetization (center panels) already 
hint at the contrast expected with a magnetic STM tip which is 
sensitive to a magnetization component given by its own 
magnetization direction. We find that the maxima and minima need 
not correspond to the atom positions and that the images can 
change qualitatively with the tip magnetization. A further striking 
point is evident from the local magnetization (top panel), i.e.~the
rotation of the magnetization density is not uniform. Above the atom 
sites the rotation is slow and in between it is much faster. 

  \begin{figure}
  \begin{center}
  \includegraphics[width=8.6cm]{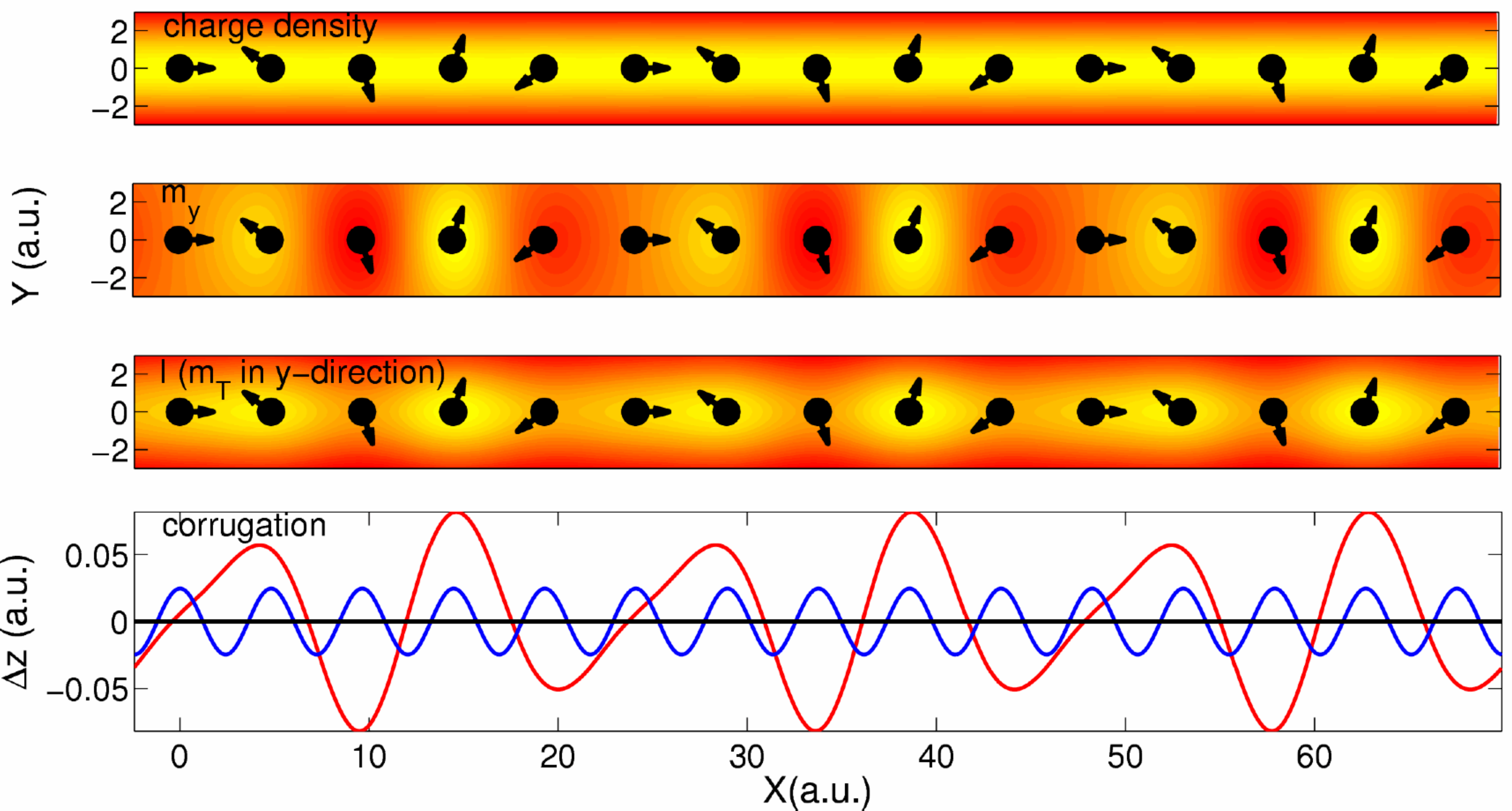}
  \caption{\footnotesize{
    (Color online) 
    Calculated SP-STM images for a Mn chain on Cu(110) in a 
    spin-spiral state with $q=0.4\times \frac{2\pi}{a}$. A height of 
    $z=5.9\, \text{\AA}$, a bias voltage of $-0.3$~V, and a tip 
    spin-polarization of 0.4 have been chosen. The tip magnetization 
    direction is assumed along the $y$-direction. In the upper three 
    panels the black spheres show the Mn atoms and the arrows 
    indicate their spin moments which show the magnetic periodicity 
    of 5 atoms. The upper two panels display the non-spin-polarized 
    and the spin-polarized component of the tunneling current, 
    respectively. The third panel depicts the resulting total tunneling 
    current. The bottom panel shows line scans along the chain axis 
    obtained by simulating the constant current mode. The red line 
    corresponds to the corrugation in the spin-polarized case. The 
    corrugation for the non-spin-polarized case (blue line) is given as 
    a reference for the atom positions and has been enhanced by a 
    factor of 25.
  }} \label{current_y_pic}
  \vspace{0.1cm}
  \includegraphics[width=8.6cm]{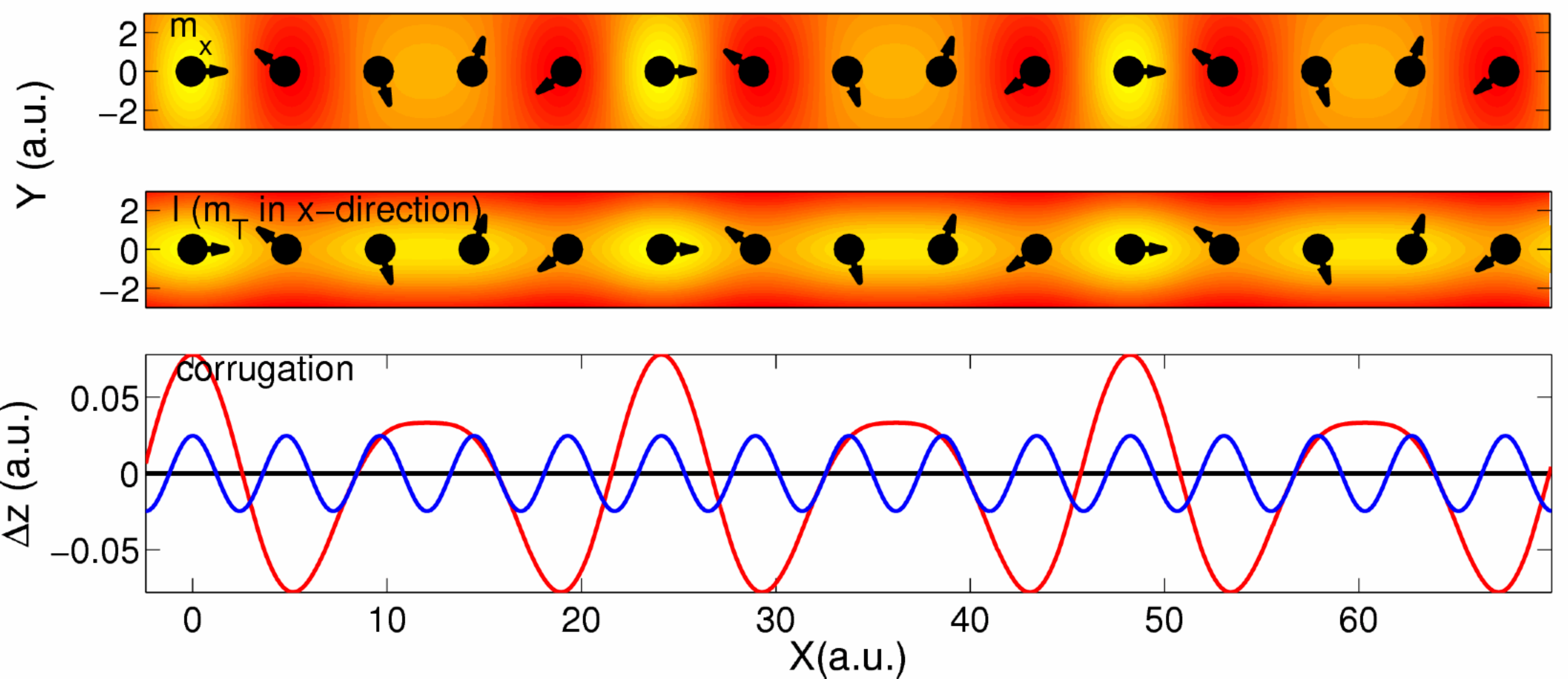}
  \caption{\footnotesize{
    (Color online) 
    Analogous to Fig.~\ref{current_y_pic}, but with the tip 
    magnetization direction pointing along the $x$-direction.
  }} \label{current_x_pic}
  \end{center}
  \end{figure}

Another interesting feature which can be seen in the upper plot of 
Fig.~\ref{dens_0.4_pic} is that the direction of the magnetization in
the vacuum coincides with that inside the Mn atoms. This appears somewhat
counterintuitive, since, according to the density of states of the Mn chain on 
Cu(110) presented in Fig.~\ref{Mndos}, the minority states dominate in the 
region around the Fermi level. In order to resolve this discrepancy, in 
Fig.~\ref{vacuum_dos} we plot the spin- and orbitally-resolved local DOS inside 
the Mn spheres together with the LDOS in the vacuum at a distance of 
$z=5.9$~\AA~from the Mn chain on Cu(110) in its ferromagnetic state 
(the case of the antiferromagnetic Mn chain is essentially analogous). The 
orbitally-resolved LDOS (Fig.~\ref{vacuum_dos}(a)) shows that these
are the Mn $d$-states which dominate at $E_F$ for both spin directions, and 
while for spin-down the $d$-DOS is quite large, for spin-up it is rather small 
and featureless, owing to the exchange splitting which shifts the position of 
the spin-up $d$-states deep below the Fermi level. On the other hand, the 
situation with the LDOS in the vacuum is completely opposite, as can be seen 
from Fig.~\ref{vacuum_dos}(b). For minority spin, the vacuum LDOS around 
$E_F$ is very small, increasing drastically in value only above 1 eV. However, 
the vacuum LDOS of spin-up electrons is very large and dominates the total 
LDOS in a large window around $E_F$ and determines the direction of the local 
magnetization DOS in the vacuum above the Mn atoms, presented in 
Fig.~\ref{dens_0.4_pic}. 

By comparing  Figs.~\ref{vacuum_dos}(a) and (b) we can see that the  structure 
of the spin-up vacuum LDOS in the energy window [$-2$~eV,\,$+1$~eV] is in 
very nice correspondence to the $s$-DOS of the Mn atom. In particular, the peaks 
in the $s$-DOS at around $-1.8$~eV, $-1.1$~eV, $+0.25$~eV, $+0.5$~eV and the 
peak directly at the Fermi energy can be clearly seen also in the majority vacuum 
LDOS, while the $d$-DOS is quite featureless for energies above $-1$~eV in this 
spin channel. This suggests, that the magnetization in the vacuum high above the 
chain is actually driven by spin-polarized $s$- and $p$-electrons which reach 
further into the vacuum, rather than localized $d$-states. The spin-polarization 
of the $s$- and $p$-states stems from their hybridization with the $d$-electrons 
in the chain, which influence their bonding properties. While the majority 
$s$- and $p$-states bare an anti-bonding character with respect to the interaction
with the substrate, the minority $s$- and $p$-states are bonded stronger to the 
Cu(110) surface, which makes their spread into the vacuum much smaller.
The latter scenario has been also recently observed and explained from 
first-principles for deposited magnetic adatoms and their 
clusters\cite{Serrate,Paolo:2010} as originating from the 
reduced coordination and symmetry of the adsorbates. This explains why the 
localized Mn spin moment due to spin-up states well below the Fermi energy 
is collinear with the magnetization density in the vacuum, obtained by integrating 
the states in a small window around $E_F$. Such a correspondence between the 
direction of the atomic spin moment in the deposited chain and the vacuum 
magnetization was also found for biatomic Fe chains on an Ir 
substrate.\cite{MokrousovFeIr}

Now we turn to the calculated spin-polarized STM images which have 
been obtained applying the spin-polarized version of the 
Tersoff-Hamann model~\cite{D_Wortmann} and using the electronic 
structure calculated within DFT. Figure~\ref{current_y_pic} displays 
the result for a tip magnetization along the $y$-direction. The upper 
two panels show the non-spin-polarized and the spin-polarized 
component of the tunneling current, respectively, at a fixed distance 
from the chain atoms of 5.9~{\AA}. As expected the atom positions 
are not resolved in the non-spin-polarized part of the current within 
the Tersoff-Hamann approximation. The apparent width of the chain 
is about 2~{\AA}. The spin-polarized part, on the other hand, clearly 
shows the five atom periodicity of the spin spiral structure with 
$q=0.4\times \frac{2\pi}{a}$ which implies an angle of 
$\alpha=144^\circ$ between the spins of neighboring atoms. In the 
total tunneling current, the superstructure due to the magnetic signal 
is still visible, but it is slightly damped due to the large contribution 
from the non-spin-polarized part. Therefore, the spin structure should 
be resolvable within an SP-STM experiment.

In order to quantify the contrast we have calculated the corrugation 
amplitude by simulating a line scan along the chain axis in the 
constant-current mode. We have chosen the starting point of the scan 
lines at a height of 5.9~{\AA} which fixes the constant current. The
obtained scan lines, shown in the bottom panel of 
Fig.~\ref{current_y_pic}, reveal an asymmetric profile and the 
maxima and minima of the spin-polarized STM image do not 
necessarily coincide with atom positions. We find a corrugation 
amplitude, i.e.~the maximum height change as the tip moves along 
the chain, of about 0.08~{\AA}. This value is above the resolution 
limit of STM (cf.~measurements of spin spirals in 
Ref.~[\onlinecite{chiralmagn}]) and is enhanced at smaller tip-sample 
distance and larger tip spin-polarization (a value of 0.4 has been 
assumed here).

In Fig.~\ref{current_x_pic} the predicted SP-STM images are 
displayed for a tip magnetization along the $x$-axis. The five atom 
periodicity is again apparent from the plot and a three-fold contrast 
pattern is also observed. The main difference between a tip 
magnetization, $\textbf{m}_T$, pointing into the $x$- and 
$y$-direction, respectively, is that there are atoms whose magnetic 
moments point directly into the $x$-direction, but there are 
no atoms whose magnetic moments directly point into the 
$y$-direction. At some points the maximum of the spin-polarized 
corrugation is therefore directly localized above the atom sites in 
the case of $\textbf{m}_T$ pointing into the $x$-direction, while 
when $\textbf{m}_T$ pointing into the $y$-direction this is never 
the case. For this reason, the two schemes are not only shifted with 
respect to each other, but look substantially different. In 
Fig.~\ref{current_x_pic} the third and the fourth atom are equivalent
in the sense that their magnetizations have the same $\alpha$
component and therefore they produce the same contrast. 
The same argument holds for atom 2 and 5. On the contrary, in the 
case of $\textbf{m}_T$ pointing in the $y$-direction these atoms 
are not equivalent.

Interestingly, the vacuum magnetization can be well reproduced 
within a simple model of SP-STM,~\cite{independent_orbital_approx} 
which relies only on a knowledge of the magnetic structure and 
assumes a spherical exponential decay of the wave functions. Therefore, 
the discussed SP-STM images could have also been obtained within this 
much faster approach. Simulations for spin spirals with a different pitch, 
i.e.~other values of $q$, are therefore easily possible in this way.

\section{Summary and Outlook}

We have performed an {\it ab initio} study of the non-collinear
magnetism in freestanding and deposited monatomic Mn chains
neglecting the effect of spin-orbit coupling, which is relatively
small in our systems. We find that the spin-spiral magnetic state
is the ground state of unsupported Mn chains in a large interval
of distance between the atoms. We were able to attribute the
appearance of non-collinear magnetism in this system to the
existence of a ferromagnetic-to-antiferromagnetic transition
point. In the vicinity of this point the nearest-neighbor exchange
parameter changes sign and is on the order of 
exchange interactions beyond nearest neighbors, which promotes
an energy minimum of a spin-spiral state. We also demonstrate,
within the virtual crystal approximation, that upon alloying chains
of Mn with Cr and Fe the ground state of Mn chains can be
controlled. In particular, while Cr-rich Mn chains are
antiferromagnets, upon increasing the Mn concentration they
exhibit a spin spiral ground state. Upon mixing Mn chains with Fe
a transition from a non-collinear ground state to the FM state
occurs. The trajectory which Mn chains take in the magnetic phase
space upon alloying with Cr and Fe is therefore smooth.

For Mn chains on the (110) surfaces of fcc Cu, Pd and Ag we
observe an overall strong tendency towards antiferromagnetism.
As a result, among the considered substrates only Cu(110) and
Ag(110) allow for a spin-spiral ground state in the deposited Mn 
chains close to the AFM solution with a modest gain in energy 
of 3.5 and 5.5~meV over the AFM state, respectively. We 
attribute this suppression of the tendency of Mn
chains towards non-collinear magnetism, which is extremely
pronounced for unsupported chains, to the effect of the hybridization
with the substrate. Within the Alexander-Anderson model of
exchange between $d$-states it can be shown that the hybridization
weakens the ferromagnetic double exchange mechanism and the
antiferromagnetic kinetic exchange interaction prevails.\cite{Phivos}

An effect which can enhance the spin-spiral minimum of Mn chains on
heavier substrates lies in the spin-orbit driven Dzyaloshinskii-Moriya 
interaction (DMI).\cite{chiralmagn} Irrespective of its sign, adding the 
DMI contribution to the spin-spiral dispersion curve would make the 
spin-spiral minimum in the vicinity of the collinear AFM solution more 
pronounced, owing to the anti-symmetric nature of this energy correction 
with respect to the spin-spiral vector. The DMI energy contribution 
will have to compete with another manifestation of the spin-orbit coupling
in these systems $-$ the magneto-crystalline anisotropy energy, which 
we, however, expect to be rather small for considered deposited chains.

Taking into account the stabilization of the magnetization by the weak 
spin-orbit coupling in Mn chains on the Cu and Ag substrates, we 
predict that the spin-spiral state in deposited Mn chains can be observed 
experimentally via,~e.g., SP-STM techniques. Based on the vacuum 
charge and magnetization density of supported Mn chains on Cu(110) 
we calculate SP-STM images, which can provide a conclusive evidence 
for a non-collinear ground state spin-structure in Mn chains.

Financial support of the Stifterverband f\"ur die Deutsche Wissenschaft 
is gratefully acknowledged. Y.M. thanks HGF-YIG Programme VH-NG-513 
for funding and J\"ulich Supercomputing Center for computational time. 
We would like to thank Gustav Bihlmayer and especially 
Phivos Mavropoulos and Stefan Bl\"ugel for many illuminating discussions 
and suggestions.

\end{document}